\documentclass[prd,onecolumn,preprintnumbers,amsmath,amssymb,superscriptaddress]{revtex4}
\usepackage{graphicx}% Include figure files
\usepackage{epsfig}% Include figure files
\usepackage{rotating}% Include figure files
\usepackage{bm}% bold math
\def\beq{\begin{equation}}   
\def\eeq{\end{equation}}

\def\bea{\begin{eqnarray}}
\def\eea{\end{eqnarray}}
\begin{document}

\title{Comment on $\lq\lq$New constraints of a light CP-odd Higgs
boson and related NMSSM Ideal Higgs Scenarios'' by Dermisek and 
Gunion (arXiv:1002.1971 [hep-ph])}

\author{Miguel-Angel Sanchis-Lozano}
\affiliation{Instituto de F\'{\i}sica Corpuscular (IFIC)
and Departamento de F\'{\i}sica Te\'orica, 
Centro Mixto Universitat de Val\`encia-CSIC, 
Dr. Moliner 50, E-46100 Burjassot, Valencia, Spain}

\begin{abstract} 
In two recent papers \cite{Dermisek:2010mg,Dermisek:2009fd}  
Dermisek and Gunion
provide new constraints on a light CP-odd Higgs boson 
in the framework of the $\lq\lq$ideal'' NMSSM (and related scenarios) based
on experimental data from LEP, CLEO, BaBar and CDF experiments.
In this brief comment we argue that special care is still
needed inside a narrow
mass window where mixing of a pseudocalar Higgs-like particle
with $\eta_b$ resonances below $B\bar{B}$
can occur. We also stress that observables
testing lepton universality and a possible
distorsion of the bottomonium mass spectrum can
provide an alternative analysis at (Super) B-factories
in the search of such an elusive light pseudoscalar Higgs-like object.

\end{abstract}

\maketitle

Recent measurements by BaBar \cite{Aubert:2009cka}, CLEO \cite{Love:2008hs}, 
ALEPH \cite{Collaboration:2010aw}
and CDF \cite{Aaltonen:2009rq} 
have allowed the authors of \cite{Dermisek:2010mg,Dermisek:2009fd} 
to provide new and stringent constraints on a 
light CP-odd Higgs boson (denoted here as $A$)
coupling to down-type fermions
in the framework of the NMSSM (or similar models). However, 
a caveat is in order inside a narrow mass window where
$A-\eta_b$ mixing should occur \cite{Drees:1989du,Fullana:2007uq}, 
ultimately
resulting in a negative influence on the experimental detection of
a new state typically expected to show up as a single peak
in the invariant mass spectrum, because:

\begin{itemize}

\item[{\it i)}] The total width of the physical (mixed) CP-odd Higgs state
could substantially increase since the $\eta_b$ resonance(s)
would have total width(s) of ${\cal O}(10)$ MeV, not negligible
compared to experimental resolution as usually assumed in the experimental
searches. Actually, since we are dealing with mixed states, what should be
understood as pseudoscalar Higgs state is, to some extent, 
a matter of convention. It seems natural to call $\lq\lq$Higgs'' 
the mass eigenstate with the largest 
$A$-component ($P_{i,4}$) of all four possible mixed states
($\eta_i$, $i=1,2,3,4$):
\[
\eta_i = P_{i,1}\;\eta_b^0(1S)
+ P_{i,2}\;\eta_b^0(2S)+ P_{i,3}\;\eta_b^0(3S) + P_{i,4}\;A
\]
where $\eta_{b}^0(nS)$ and $A$ denote the unmixed states; $P_{i,4}$
varies as a function of $m_A$ as can be seen from the middle plot of Fig.1. 
The resulting mass spectrum is shown in the left-hand plot of Fig.1
(see \cite{Domingo:2009tb} for more details). 

\item[{\em ii)}] Production and decay into leptons of a CP-odd Higgs 
would be channeled through distinct physical particles
with different masses. Therefore, a multi-peak scenario would 
show up instead of a single narrow peak,  
whenever a significant mixing occurs,  
in either the photon-energy spectrum (from radiative
Upsilon decays at B factories), or the dimuon mass spectrum
(at hadron colliders).

\end{itemize}

%\captionsetup{aboveskip=-6pt,belowskip=-8pt}
\begin{figure}[ht!]
\begin{center}
\includegraphics*[width=0.31\linewidth,height=0.31\linewidth]
{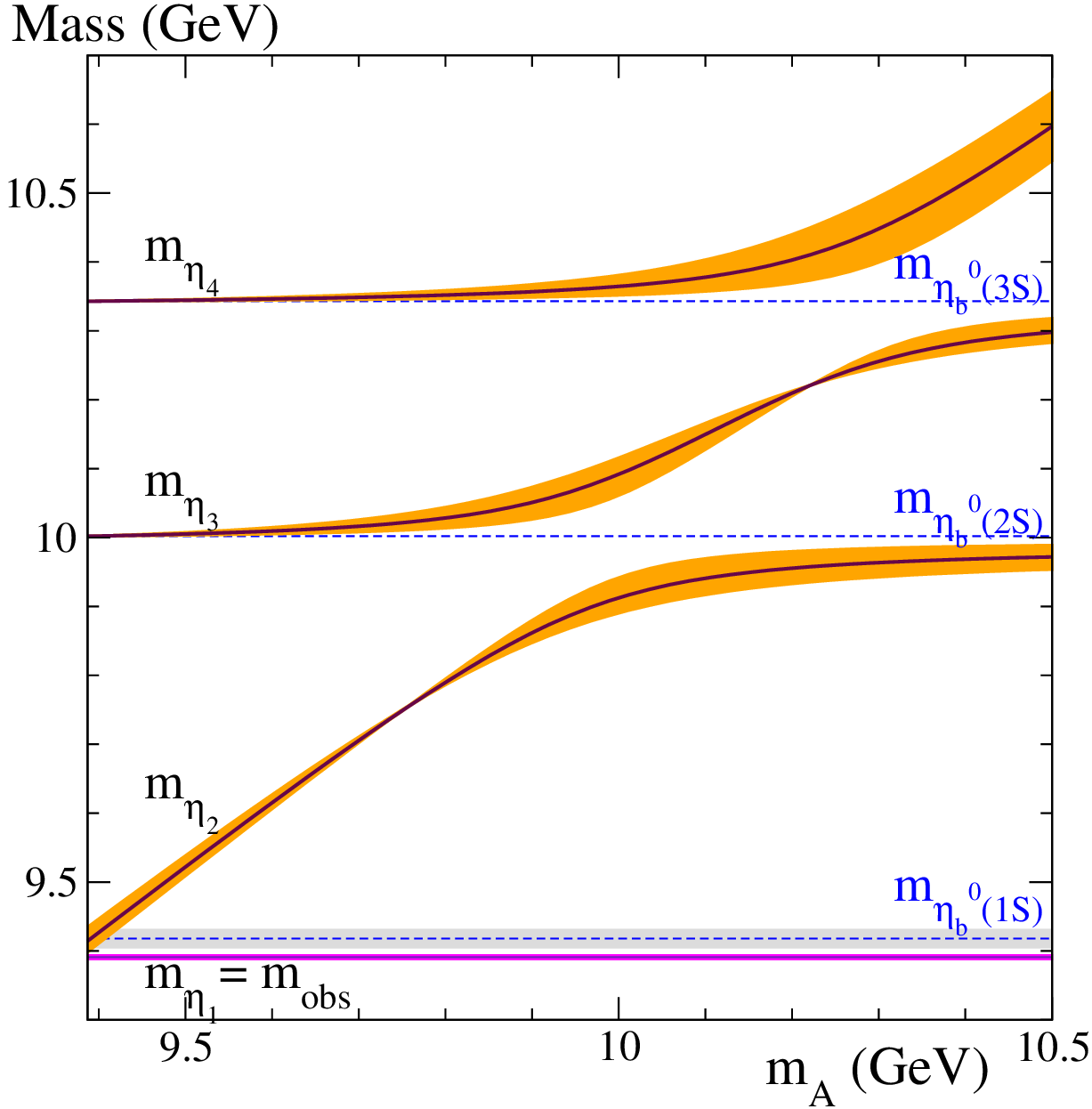}
\includegraphics*[width=0.31\linewidth,height=0.31\linewidth]
{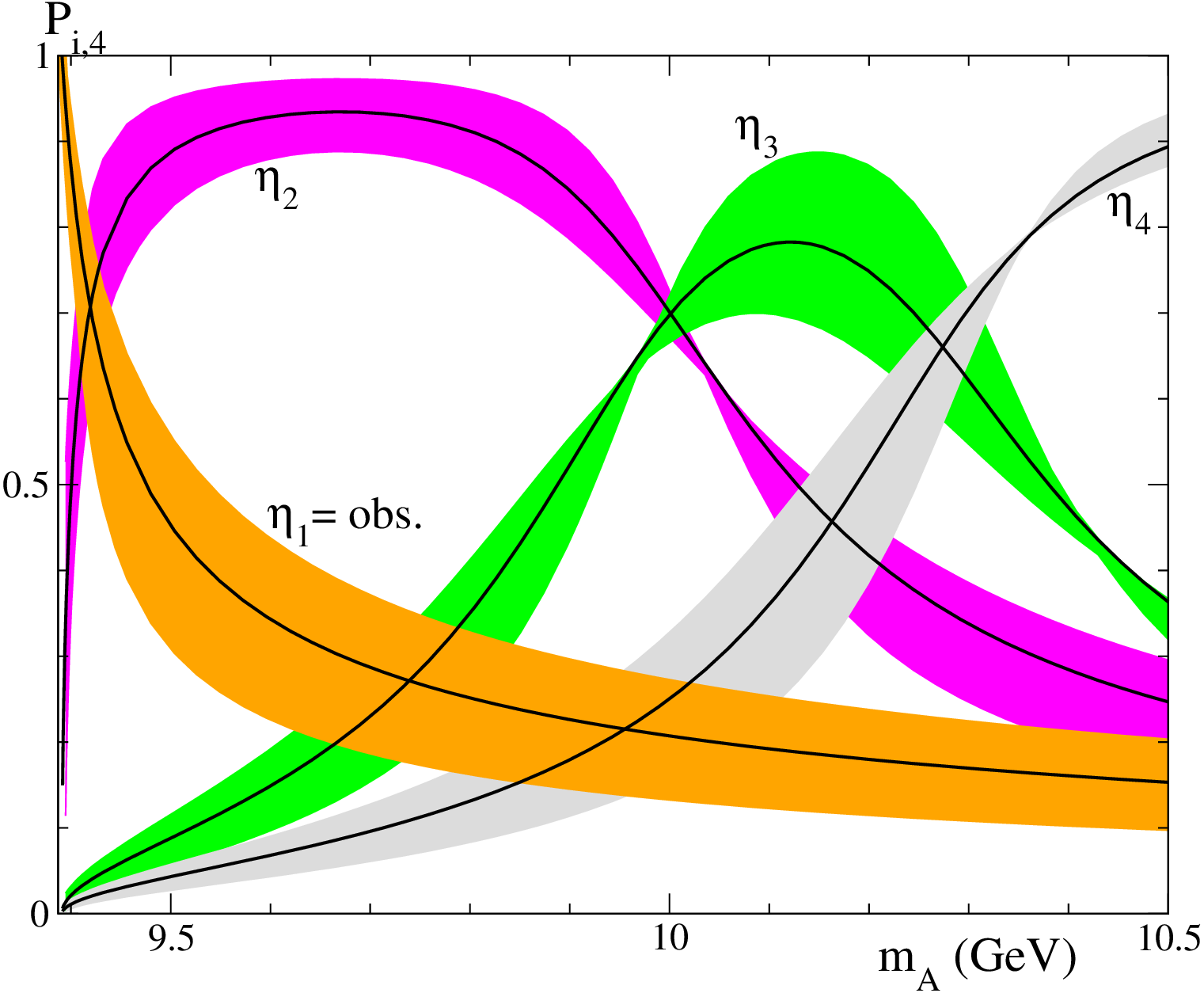}
\includegraphics*[width=0.31\linewidth,height=0.31\linewidth]
{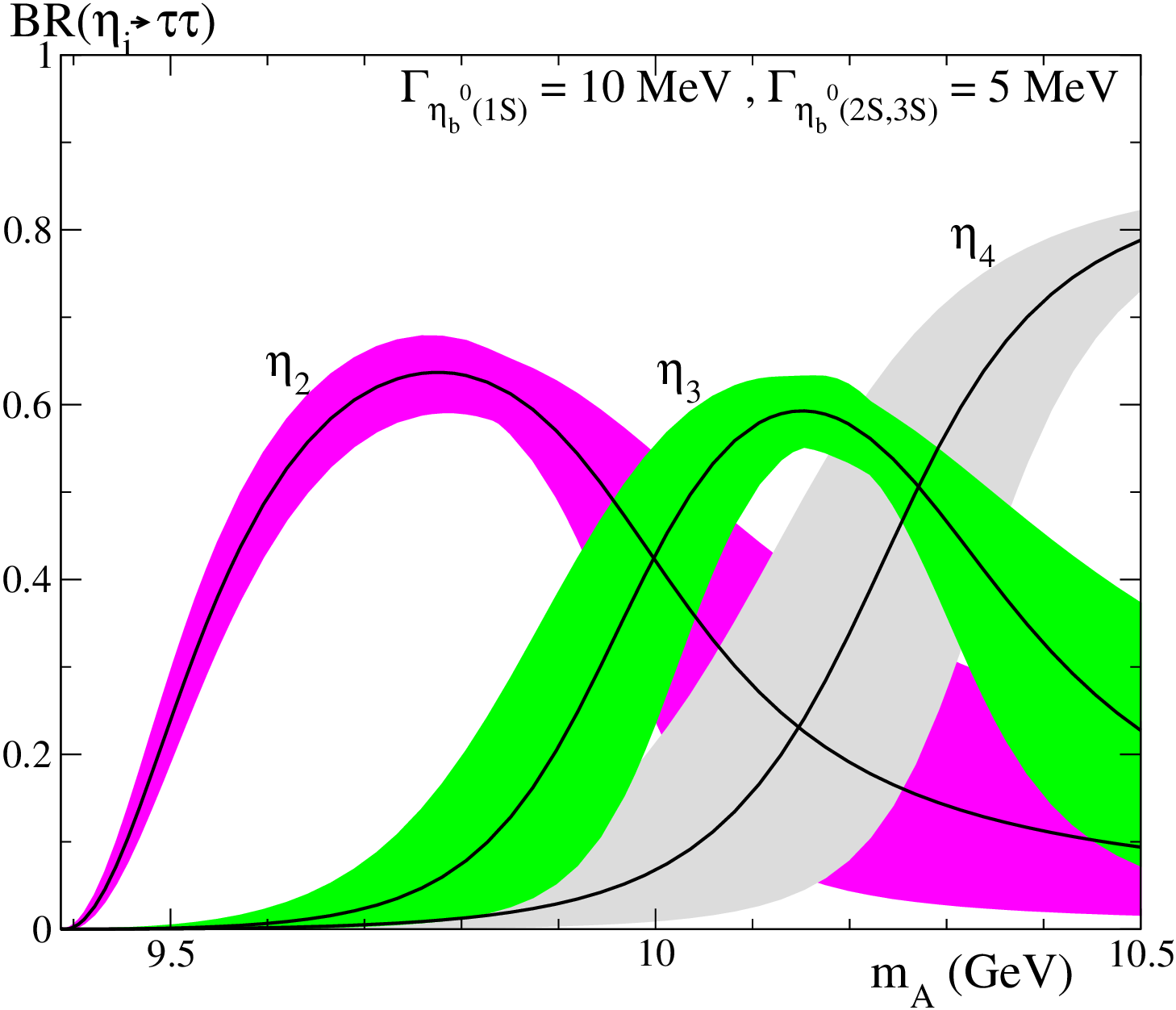}
\end{center}
\addvspace{-6mm}
\caption{{\em Left}: Masses of the physical (mixed) pseudoscalar 
states ($\eta_{1,2,3,4}$) below $B\bar{B}$ threshold as
function of the unmixed $A$ mass obtained in \cite{Domingo:2009tb}
by requiring that the difference between the 
perturbative QCD expectation
and the measured  $\eta_b(1S)$ mass \cite{:2008vj,Bonvicini:2009hs}
is entirely ascribed to the $A-\eta_b(1S)$ mixing. 
{\em Middle}: The $A$-component $P_{i,4}$
of all 4 eigenstates versus $m_A$. {\em Right}: 
Tauonic branching ratios of $\eta_{2,3,4}$ eigenstates versus $m_A$;  
$BR(\eta_1\to \tau^+\tau^-)< 8\%$ \cite{Aubert:2009cka} is 
not shown in the plot.  Solid (dashed) lines stand
for the (un)mixed states and colored fringes 
indicate theoretical uncertainties \cite{Domingo:2009tb}.}
\end{figure}

For example, the $\Upsilon(3S)\to\gamma\tau^+\tau^-$ decay rate
via the new physics contribution
would be significantly distributed among different channels
(i.e. through intermediate $\eta_{2,3,4}$ states) as $m_A$ varies along
the $[9.4,10.5]$ GeV range (see the right-hand plot of Fig.1), leading 
to weaker individual signals than expected. Moreover, let us 
mention that the Wilczek formula for $\Upsilon \to \gamma A$ decays
becomes unreliable to set exclusion limits 
above $m_A \simeq 9$ GeV, 
because of large theoretical uncertainties due to bound state, 
QCD, and relativistic corrections \cite{guide,Domingo:2009tb}.

A similar argument related to the spreading of any light Higgs
signal would equally apply to searches in the dimuon mass
spectrum measured by CDF \cite{Aaltonen:2009rq}, despite the fact that
the production mechanism (via quark-loop induced $ggA$ coupling)
of $\eta_i$ states is different from the previous case. In addition,
experimental smearing would likely lead to bumps rather than
well-separated peaks in the mass spectrum under study.
Therefore the constraints obtained in \cite{Dermisek:2010mg,Dermisek:2009fd} 
for a CP-odd Higgs with 
$9.4 \lesssim m_A \lesssim 10.5$ GeV should still be taken with care,
not definitely excluding 
larger couplings to down-type fermions 
accounting for the muon $g-2$ anomaly (see e.g. 
\cite{Gunion:2008dg,Domingo:2007dx} and references therein).

Let us finally comment on the LEP result recently reported by
ALEPH \cite{Collaboration:2010aw}
on the production and non-standard decay of a Higgs boson $h$
into four taus through intermediate pseudoscalars $A$, 
where exclusion limits are set for a combined production
cross section times branching ratios, namely
\[
\xi^2=\frac{\sigma(e^+e^-\to Zh)}{\sigma_{SM}(e^+e^- \to Zh)} \times\
BR(h\to AA)\ \times\ BR(A \to \tau^+\tau^-)^2
\]
Notice that the above expression lacks exact physical meaning under the 
hypothesis of $A-\eta_b$ mixing, for $A$ could not be
a (single) on-shell state anymore (as likely assumed in 
the ALEPH analysis) but a component of 
different $\eta_i$ eigenstates as already argued before.
Certainly, setting experimental limits on 
a quantity like $\xi^2$ as done by ALEPH 
is definitely useful in the hunt for a light
Higgs boson, but requires a reinterpretation of the
factor $BR(A \to \tau^+\tau^-)$ in terms of a set of $\eta_i$
intermediate states in the subsequent analysis
(e.g. observed versus expected upper limits). Let us
recall, in this regard, the search for Yukawa production
at LEP of a light neutral Higgs boson carried out by
OPAL \cite{Abbiendi:2001kp}, 
where mixing with $b\bar{b}$ bound states
was taken into account in the data analysis (modifying the branching
ratios into taus accordingly \cite{Drees:1989du}), 
implying considerably looser bounds for the 
pseudoscalar Higgs coupling to 
down-type fermions, in this mass range.

On the other hand, 
observables based on inclusive measurements, e.g. testing
lepton universality in $\Upsilon$ decays (i.e. all leptonic
branching ratios have to coincide aside lepton mass effects)
\cite{Domingo:2008rr,Fullana:2007uq,SanchisLozano:2003ha,SanchisLozano:2002pm}
could provide an alternative way to determine exclusion limits 
for a light pseudoscalar Higgs. In fact, a
recent result from BaBar 
in $\Upsilon(1S)$ decays finds no significant
deviation from the SM expectation \cite{:2010bt}. Let us
emphasize, however, that lepton universality breaking should become
experimentally sizeable for $\Upsilon(2S)$ and $\Upsilon(3S)$ 
decays as pointed out in \cite{Domingo:2008rr};
thereby we strongly suggest the BaBar Collaboration 
extend their analysis to the two latter cases.

Finally, let us stress that a possible
distorsion of the $\eta_b$ mass levels \cite{Domingo:2009tb}, 
as shown in the left-hand plot of Fig.1 could become another
interesting way of seeking a light CP-odd Higgs
in the range $[9.4,10.5]$ GeV.
Although this searching strategy is
free of the above-mentioned theoretical uncertainties
plaguing the Wilczek formula, 
any new physics signal
manifesting as unexpectedly large or small (even negative!)
hyperfine splittings ($m_{\Upsilon(nS)}-m_{\eta_b(nS)}$)
requires a good control of both perturbative and non-perturbative
(e.g. lattice) QCD calculations of the bottomonium system 
\cite{Brambilla:2004wf,Gray:2005ur,Penin:2009wf}.

\end{document}